\documentclass[final,onefignum,onetabnum]{siamart171218}
\usepackage{graphicx}
\usepackage{epstopdf}
\usepackage{bm}
\usepackage{amssymb}
\usepackage{psfrag}
\usepackage{color}
\usepackage{amsbsy}      
\usepackage{amsmath}
\usepackage{mathtools}
\usepackage{dsfont}
\usepackage{lineno}
\usepackage{comment}
\usepackage{hyperref}
\usepackage{hhline}
\usepackage{tcolorbox}
\usepackage{makecell}
\usepackage{booktabs}
\usepackage{float,lscape}
\usepackage{dsfont}
\usepackage{url}
\usepackage{braket}
\usepackage{etoolbox}
\usepackage{soul, color}
\usepackage{amsopn}
\def\ie{\textit{i.e.}}
\def\eg{\textit{e.g.}}
\graphicspath{{figs/}}

\newcommand{\iu}{\mathrm{i}\mkern1mu}

\ifpdf
\hypersetup{
  pdftitle={Classical and quantum random-walk centrality measures in multilayer networks},
  pdfauthor={Lucas B\"ottcher and Mason A.~Porter}
}
\fi

\usepackage{lipsum}
\usepackage{amsfonts}
\usepackage{graphicx}
\usepackage{epstopdf}
\usepackage{algorithmic}
\ifpdf
  \DeclareGraphicsExtensions{.eps,.pdf,.png,.jpg}
\else
  \DeclareGraphicsExtensions{.eps}
\fi

\newsiamremark{remark}{Remark}
\newsiamremark{hypothesis}{Hypothesis}
\crefname{hypothesis}{Hypothesis}{Hypotheses}
\newsiamthm{claim}{Claim}

\headers{Classical and quantum random-walk centrality measures}{L. B\"ottcher and M. A. Porter}

\title{Classical and quantum random-walk centrality measures in multilayer networks\thanks{Submitted to the editors on December 13, 2020.
\funding{LB received funding from the Swiss National Fund (P2EZP2\_191888) and from the Army Research Office (W911NF-18-1-0345).}}}

\author{Lucas B\"ottcher\thanks{Dept.~of Computational Medicine, University of California, Los Angeles, CA, 90095-1766, United States of America and Computational Social Science, Frankfurt School of Finance and Management, Frankfurt am Main, 60322, Germany
  (\email{lucasb@g.ucla.edu}).}
\and Mason A.~Porter\thanks{Dept.~of Mathematics, University of California, Los Angeles, CA, 90095-1766, United States of America 
  (\email{mason@math.ucla.edu}).}}

\usepackage{amsopn}

\makeatletter
\newcommand*{\addFileDependency}[1]{
  \typeout{(#1)}
  \@addtofilelist{#1}
  \IfFileExists{#1}{}{\typeout{No file #1.}}
}
\makeatother

\begin{document}

\maketitle

\begin{abstract}
Multilayer network analysis is a useful approach for studying the structural properties of entities with diverse, multitudinous relations. Classifying the importance of nodes and node-layer tuples is an important aspect of the study of multilayer networks. To do this, it is common to calculate various centrality measures, which allow one to rank nodes and node-layers according to a variety of structural features. In this paper, we formulate occupation, PageRank, betweenness, and closeness centralities in terms of node-occupation properties of different types of continuous-time classical and quantum random walks on multilayer networks. We apply our framework to a variety of synthetic and real-world multilayer networks, and we identify marked differences between classical and quantum centrality measures. Our computations also give insights into the correlations between certain random-walk-based and geodesic-path-based centralities.
\end{abstract}



\section{Introduction}
Centrality measures~\cite{baek2019social} are useful quantities to rank nodes in a network according to their importances~\cite{newman2018networks}. Depending on the centrality measure that one calculates, highly central nodes may be ones that have the largest numbers of neighbors (degree centrality), lie on many shortest paths between two nodes (geodesic betweenness centrality), have a small distance (\eg, via a sum of shortest paths) to all other nodes (closeness centrality), and so on. These and other centrality measures have a large variety of applications, including identifying important spreaders of diseases or information~\cite{shah2011rumors,luo2013identifying}, ranking websites and other objects~\cite{gleich2015pagerank}, and characterizing granular and particulate structures~\cite{papadopoulos2018network,nauer2019random}.

A variety of centrality measures have been developed for monolayer networks and generalized to multilayer networks~\cite{kivela2014multilayer,aleta2019multilayer,taylor2019tunable}. One possible formulation of centrality in monolayer and multilayer networks is based on the node-sampling properties of different types of classical random walks (CRWs)~\cite{noh2004random,sola2013eigenvector,sole2013spectral,de2015ranking,sole2016random,masuda2017random}. 
In parallel to the recent developments in network analysis, the study of quantum versions of CRWs called \emph{continuous-time quantum walks} (CTQWs)~\cite{childs2004spatial,portugal2013quantum} has led to several insights into the influence of quantum effects on the propagation properties of random walks on networks. Quantum walks have been implemented in various experimental settings, including nuclear-magnetic-resonance setups~\cite{ryan2005experimental}, trapped neutral-atom~\cite{karski2009quantum} and trapped ion~\cite{Schmi09,Zaeh10} systems, and photonic systems~\cite{bouwmeester1999optical,Do05,Pere08,Schr10}. One appealing property of quantum walks (QWs) is that they propagate quadratically faster than their classical counterparts on certain networks~\cite{childs2004spatial}. One can achieve full quantum speed-up on many networks, including regular graphs~\cite{janmark2014global}, Erd\H{o}s--R\'{e}nyi (ER) networks~\cite{chakraborty2016spatial}, and $d$-dimensional cubic periodic lattices with $d \geq 5$~\cite{childs2004spatial}. The advantage of quantum search strategies has been termed the ``Grover speed-up'' in recognition of Lov Grover's foundational work on this topic~\cite{grover1996fast,grover1997quantum}. Recent studies~\cite{berry2010quantum,mahasinghe2014quantum} have analyzed the connection between the run time and success probability (\ie, the probability of finding a quantum walker at a target node at a certain time) of quantum-search algorithms and the centrality of a ``marked'' (\ie, target) node. For discrete-time quantum walks (DTQWs), it has been shown that the success probability of a marked node does not necessarily increase with its closeness centrality~\cite{berry2010quantum}. However, for Cayley trees, the success probability is large for nodes with a small ``eccentricity''~\cite{mahasinghe2014quantum}, which is defined as the maximum distance from a marked node to all other nodes. A later study~\cite{philipp2016continuous} illustrated that the run time of CTQW-based search algorithms on balanced trees is correlated with the closeness centrality of a marked node. Notably, all of these works computed centrality measures using classical algorithms. However, as described in Refs.~\cite{sanchez2012quantum,rossi2014node}, it is also possible to define centrality measures that are based on the node-occupation statistics of QWs. Subsequently, numerical and experimental implementations of CTQW-based occupation centralities were compared to several classical centrality measures on random networks~\cite{izaac2017centrality}. Very recently, it was shown that one can tune node-occupation statistics of CRWs and CTQWs by using appropriate stochastic resetting protocols~\cite{wald2020classical}.

In the present paper, we use the framework of Ref.~\cite{sole2016random} and compare several classical and quantum random-walk-based centrality measures on multilayer networks. Our paper proceeds as follows. In Sec.~\ref{sec:multilayer}, we give an overview of the mathematical formulation of multilayer networks. We then derive the evolution equations of CRWs and CTQWs on multilayer networks in Sec.~\ref{sec:cl_q_walks}. In Sec.~\ref{sec:centralities}, we use these evolution equations to define classical and quantum versions of random-walk-occupation centrality, PageRank centrality, betweenness centrality, and closeness centrality. In Sec.~\ref{sec:numerics}, we calculate these centrality measures for a variety of synthetic and empirical multilayer networks. We also examine correlations between (1) random-walk and geodesic versions of betweenness centrality and (2) between random-walk and geodesic versions of closeness centralities. We conclude our study in Sec.~\ref{sec:conclusion}. Our source code (and additional information on parallelization methods) is publicly available at~\cite{GitLab}.
%
%
\section{Multilayer networks}
\label{sec:multilayer}
\begin{figure}
\centering
\includegraphics[width=0.85\textwidth]{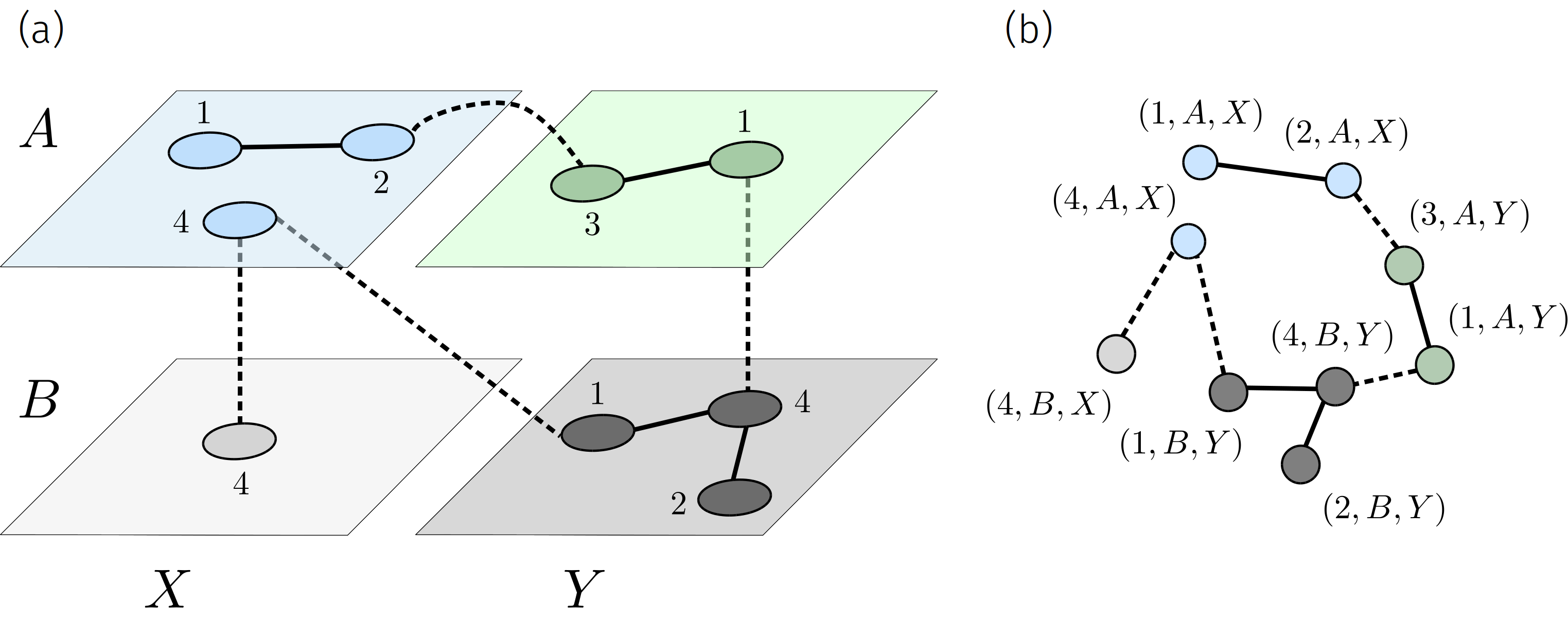}
\caption{\textbf{Example of a multilayer network}. (a) We show a multilayer network $M=(V_M,E_M,V,L)$ with four nodes (\ie, $V=\{1,2,3,4\}$) and two aspects. These two aspects have corresponding elementary layers $L_1=\{\mathrm{A},\mathrm{B}\}$ and $L_2=\{\mathrm{X},\mathrm{Y}\}$. The resulting four layers of $M$ are $\{\mathrm{A},\mathrm{X}\}$, $\{\mathrm{A},\mathrm{Y}\}$, $\{\mathrm{B},\mathrm{X}\}$, and $\{\mathrm{B},\mathrm{Y}\}$. Not all nodes are present in all layers. In the depicted example, the set $V_M$ of node-layer tuples is $V_M=\{(1,\mathrm{A},\mathrm{X}),(2,\mathrm{A},\mathrm{X}),(4,\mathrm{A},\mathrm{X}),(4,\mathrm{B},\mathrm{X}),(1,\mathrm{A},\mathrm{Y}),(3,\mathrm{A},\mathrm{Y}),(1,\mathrm{B},\mathrm{Y}),(2,\mathrm{B},\mathrm{Y}),(4,\mathrm{B},\mathrm{Y})\}\subseteq V\times L_1\times L_2$. (b) The graph $G_M=(V_M,E_M)$ that corresponds to the multilayer network $M$. We indicate intralayer and interlayer edges using solid and dashed arcs, respectively.
}
\label{fig:multilayer}
\end{figure}
A graph (\ie, a monolayer network) $G = (V,E)$ is an ordered pair $(V,E)$, where $V$ is a set of $N$ nodes and $E \subseteq V \times V$ is a set of edges that connect pairs of nodes. We label the nodes with the numbers $1,2,\ldots,N$. For each node $v \in V$, the degree $\text{deg}(v)$ is the number of edges that are attached to $v$.

To allow more than one layer in a network (see Fig.~\ref{fig:multilayer}), we use the multilayer-network formalism of Refs.~\cite{de2013mathematical,kivela2014multilayer}. A multilayer network $M$ has not only nodes and edges, but also layers with $d$ distinct aspects (\ie, types of layering, such as multiple types of relationships or multiple communication patterns) \cite{porter2018multilayer}. Each aspect $a$ has a corresponding \emph{elementary layer} $L_a$. We denote the sequence of sets of elementary layers by $L = \{L_a\}_{a=1}^d$. All possible combinations of elementary layers are given by the Cartesian product $L_1 \times \dots \times L_d$, and elements of this set are called ``layers''. We denote the number of layers in $M$ by $K$. 

Not all nodes need to be present in all $K$ layers. We use $V_M \subseteq V \times L$ to denote the set node-layer tuples $(v, l) \in V_M$ in which node $v$ (which represents some entity) is present in layer $l \subseteq L_1 \times \dots \times L_d$. We denote the total number of nodes in $M$ by $N=|V_M|$ and the subset of edges between node-layer tuples (\ie, ``node-layers'') by $E_M \subseteq V_M \times V_M$. The edge $((u, \alpha), (v, \beta))\in E_M$ indicates that there is an edge from node $u$ in layer $\alpha$ to node $v$ in layer $\beta$ (and vice versa, if $M$ is undirected). Each aspect of $M$ represents a type of layering, such as connections between (classical or quantum) circuits, a point in time, or some other property. We use the adjectives \emph{intralayer} and \emph{interlayer} to refer to edges between nodes within the same layer and between different layers, respectively. A multilayer network $M$ (see Fig.~\ref{fig:multilayer}) is a quadruplet $M = (V_M , E_M , V, L)$. A multiplex network is a specific type of multilayer network in which all of the interlayer edges are edges between nodes and their counterparts in other layers.
\section{Classical and quantum random walks}
\label{sec:cl_q_walks}
We consider an undirected and unweighted multilayer network $M$ with adjacency tensor $\mathcal{A}$. The adjacency-tensor components are
\begin{equation}
    \mathcal{A}^{i \alpha}_{j \beta}=
        \begin{cases}
1\,, \quad &\text{if $(i,\alpha)$ is adjacent to $(j,\beta)$} \\
0\,, \quad &\text{otherwise}\,.
        \end{cases}
\end{equation}
Note that $\mathcal{A}^{i \alpha}_{j \beta}=\mathcal{A}^{j \beta}_{i \alpha}$ for undirected networks. The transition probability of going from node $i$ in layer $\alpha$ to node $j$ in layer $\beta$ is $T^{i \alpha}_{j \beta}$. To mathematically describe a CRW on $M$, let $p_{j \beta}(t)$ be the probability that a random walker is on node $j$ in layer $\beta$ at time $t$. The corresponding temporal evolution of $p_{j \beta}(t)$ is 
\begin{align}
    p_{j \beta}(t+\Delta t)=p_{j \beta}(t)-\Delta t \left[ \sum_{(i,\alpha)\in V_{M}} T_{i \alpha}^{j \beta} p_{j \beta}(t)-\sum_{(i,\alpha)\in V_{M}} T_{j \beta}^{i \alpha} p_{i \alpha}(t)\right] \, .
\label{eq:rw1}
\end{align}
As in the description of random-walk (and hence diffusion) dynamics on a monolayer network with the combinatorial graph Laplacian, we write $T_{j \beta}^{i \alpha} = \mathcal{A}^{i \alpha}_{j \beta}/k_{i \alpha}$, where $k_{i \alpha}=\sum_{(j,\beta)\in V_{M}} \mathcal{A}^{j \beta}_{i \alpha}$ is the degree of node $i$ in layer $\alpha$~\cite{jeub2015local}. For brevity, we use Einstein summation convention and write $k_{i \alpha}=\mathcal{A}^{j \beta}_{i \alpha} u_{j \beta}$, where all components of the covariant order-2 tensor $u_{j \beta}$ are equal to $1$. Using the Einstein convention and $T_{j \beta}^{i \alpha}=\mathcal{A}^{i \alpha}_{j \beta}/k_{i \alpha}$, we rewrite Eq.~\eqref{eq:rw1} as
\begin{equation}
    p_{j \beta}(t+\Delta t)=p_{j \beta}(t)-\Delta t \left[ \delta^{i \alpha}_{j \beta} p_{i \alpha}(t)-\mathcal{A}^{k \gamma}_{j \beta} {D_{k \gamma}^{i \alpha}}^{-1} p_{i \alpha}(t) \right]\,,
\label{eq:rw2}
\end{equation}
where the Kronecker delta $\delta^{i \alpha}_{j \beta}=1$ if and only if $(i,\alpha)=(j,\beta)$ (so $\delta^{i \alpha}_{j \beta}= 0$ otherwise)
and
\begin{equation}
    D_{j \beta}^{i \alpha}=
\begin{cases}
k_{i \alpha}\,, \quad &\text{if $(i,\alpha)=(j,\beta)$}\\
0\,, \quad &\text{otherwise}
\end{cases}
\end{equation}
are the components of the degree tensor. In Eq.~\eqref{eq:rw2}, we used the fact that $\sum_{(i,\alpha)\in V_M} T^{j\beta}_{i \alpha}=1$ (\ie, that the probability of going from node $j$ in layer $\beta$ to some other node-layer is $1$, as a random walker is required to move somewhere). This multilayer formulation of random-walk dynamics naturally leads to the definition of the components~\cite{de2013mathematical} $L_{j \beta}^{i \alpha} =D_{j \beta}^{i \alpha} - \mathcal{A}^{i \alpha}_{j \beta}$ of the Laplacian tensor.

For our analytical and numerical treatments of classical and quantum random walks on multilayer networks, we apply a \emph{flattening} function~\cite{kolda2009tensor,kivela2014multilayer} to the adjacency tensor and other tensors to transform them into associated $\mathcal{N}\times \mathcal{N}$ matrices with entries in $\mathbb{R}_{\geq 0}$. The scalar $\mathcal{N}$ denotes the number of nodes in $G_M=(V_M,E_M)$, the flattened graph representation of $M$. In our paper, we refer to the elements of $V_M$ as ``node-layers'' because each node in $G_M=(V_M,E_M)$ is a node-layer of $M$. The supra-Laplacian matrix~\cite{gomez2013diffusion} $\mathbf{L}_{\rm M}=\mathbf{D}_{\rm M}-\mathbf{A}_{\rm M}$ allows us to write the evolution equation \eqref{eq:rw2} in continuous time as
\begin{equation}
    \frac{\mathrm{d} }{\mathrm{d} t} \mathbf{p} = -\mathcal{H}_{\rm c} \mathbf{p} \quad \text{with} \quad \mathcal{H}_{\rm c} = \mathbf{L}_{\rm M} \mathbf{D}_{\rm M}^{-1}\,.
\label{eq:cl_evolution}
\end{equation}
The classical Hamiltonian $\mathcal{H}_{\rm c}$ is the generator of time translation of the flattened stochastic vector $\mathbf{p}$. Each component of $\mathbf{p}$ corresponds to a node-layer. In accordance with Refs.~\cite{faccin2013degree,wald2020classical}, we use the normalized supra-Laplacian matrix $\hat{\mathbf{L}}_{\rm M} = \mathbf{D}_{\rm M}^{-1/2}\mathbf{L}_{\rm M}\mathbf{D}_{\rm M}^{-1/2}$ to formulate the evolution of a CTQW on multilayer networks according to the Schr\"odinger equation\footnote{In this equation, we set $\hbar =1$. Additionally, the choice of the quantum evolution operator is not unique. Common choices of Hermitian operators are the normalized (supra-)Laplacian matrix and the (supra-)adjacency matrix~\cite{wong2016laplacian}.}
\begin{equation}
    \frac{\mathrm{d} }{\mathrm{d} t} \ket{\psi} = - \iu \mathcal{H}_{\rm q}\ket{\psi} \quad \text{with} \quad \mathcal{H}_{\rm q} = \hat{\mathbf{L}}_{\rm M}\,.
\label{eq:q_evolution}
\end{equation}
Because of the unitary and time-reversible evolution equation for CTQWs, these walks (unlike their classical counterparts) do not approach a stationary distribution. The stochastic nature of CTQWs comes from the measurement process, rather than from the underlying dynamics~\cite{wald2020classical}. A key advantage of CTQWs over their classical counterparts is that they achieve quadratically faster propagation in certain networks~\cite{childs2004spatial}. This is a useful feature for quantum-walk-based centrality measures.
\section{Classical and quantum centralities}
\label{sec:centralities}
\subsection{Random-walk occupation centrality}
For an undirected and connected multilayer network $M$, a CRW reaches a unique stationary state $\mathbf{p}^*\in \mathbb{R}^{\mathcal{N}}$~\cite{norris1998markov} that satisfies
\begin{equation}
    \mathcal{H}_{\rm c} \mathbf{p}^* = 0\quad\text{and}\quad \sum_{\tilde{i}=1}^\mathcal{N} p_{\tilde{i}}^\ast=1\,,
\label{eq:cl_stat}
\end{equation}
where $\tilde{i}$ denotes a node-layer. We use a tilde to distinguish a node-layer $\tilde{i}$ of $G_M$ from a node $i$ of $M$.
The solution of Eq.~\eqref{eq:cl_stat} is
\begin{equation}
    p^\ast_{\tilde{i}} = \frac{k_{\tilde{i}}}{\sum_{\tilde{j}=1}^\mathcal{N} k_{\tilde{j}}}\,,
\label{eq:classical_analytic}
\end{equation}
where $k_{\tilde{i}}$ is the (total) degree of node-layer $\tilde{i}$ in flattened notation (\ie, in the graph $G_M$). Each component of the stochastic vector $\mathbf{p}^*$ corresponds to the stationary CRW occupation probability of a certain node-layer and defines a natural random-walk centrality measure for a multilayer network. For CTQWs, however, unitary time evolution does not lead to a stationary state. Instead, the long-time behavior of a QW is characterized by its long-time mean~\cite{faccin2013degree}
\begin{equation}
    q_{\tilde{j}}^* = \lim_{T\rightarrow\infty} \frac{1}{T} \int_0^T \braket{\tilde{j}|\rho(t)|\tilde{j}} \, \mathrm{d}t\,,
    \label{eq:q_ast_1}
\end{equation}
where $\mathrm{d}t$ is an infinitesimal time step, $\rho(t)=\ket{\psi(t)}\bra{\psi(t)}$ is a density operator, and $\ket{\tilde{j}} \in \mathbb{C}^{\mathcal{N}}$ is an orthonormal basis vector that satisfies
\begin{equation}
    \braket{i|j} = \delta_{i j}\,.
\end{equation}
As in the CRW, $q_{\tilde{j}}^*$ denotes the CTQW occupation probability of node-layer $\tilde{j}$. Therefore, it gives a type of random-walk centrality. 
\subsection{PageRank centrality}
A generalization of the above concept of determining random-walk occupation probabilities is to incorporate teleportation events from $(i,\alpha)$ to $(j,\beta)$ that occur with probability $1-a \in (0,1]$. In the classical setting, we replace the transition probability by~\cite{de2015ranking,sole2016random}
\begin{equation}
    T_{j \beta}^{i \alpha}=
a \mathcal{A}^{i \alpha}_{j \beta}/k_{i \alpha}+\frac{(1-a)}{N K} u^{i \alpha}_{j \beta}
\end{equation}
for an undirected and connected multilayer network $M$. All rank-4 tensor components $u^{i \alpha}_{j \beta}$ are equal to 1. In flattened notation, the resulting classical evolution equation is
\begin{equation}
    \frac{\mathrm{d} }{\mathrm{d} t} \mathbf{p}_{\text{p}} = -\mathcal{H}_{\rm c}^{\text{p}} \mathbf{p}_{\text{p}} \quad \text{with} \quad \mathcal{H}_{\rm c}^{\text{p}} = a \mathbf{L}_{\rm M} \mathbf{D}_{\rm M}^{-1}+(1-a) \left[\mathds{1}-\mathcal{N}^{-1}\mathbf{U}\right]\,,
\label{eq:cl_evolution_pr}
\end{equation}
where $\mathbf{U}$ is the matrix of $1$ entries. The PageRank analog of Eq.~\eqref{eq:q_evolution} is
\begin{equation}
    \frac{\mathrm{d} }{\mathrm{d} t} \ket{\psi_{\text{p}}} = - \iu \mathcal{H}_{\rm q}^{\text{p}}\ket{\psi_{\text{p}}} \quad \text{with} \quad \mathcal{H}_{\rm q}^{\text{p}} = a\hat{\mathbf{L}}_{\rm M}+(1-a)\left[\mathds{1}-\mathcal{N}^{-1}\mathbf{U}\right]\,.
\label{eq:q_evolution_pr}
\end{equation}
Note that $\mathcal{H}_{\rm q}^{\text{p}}$ is Hermitian. In Eqs.~\eqref{eq:cl_evolution_pr} and \eqref{eq:q_evolution_pr}, we use a version of PageRank in which a random walker can teleport to any node-layer. Other choices of teleportation protocol are also possible~\cite{gleich2015pagerank}. They lead to different PageRank Hamiltonians $\mathcal{H}_{\rm c}^{\text{p}}$ and $\mathcal{H}_{\rm q}^{\text{p}}$. 

For different teleportation probabilities $1-a$, we determine the classical PageRank centrality by solving
\begin{equation}
    \mathcal{H}_{\rm c}^{\text{p}} \mathbf{p}_{\text{p}}^* = 0\quad\text{and}\quad \sum_{\tilde{i}=1}^\mathcal{N} (\mathbf{p}_{\text{p}}^*)_{\tilde{i}}=1\,.
\label{eq:p_ast}
\end{equation}
The associated quantum PageRank centrality is
\begin{equation}
    (q_{\text{p}}^*)_{\tilde{j}} = \lim_{T\rightarrow\infty}\frac{1}{T} \int_0^T \braket{\tilde{j}|\rho_{\text{p}}(t)|\tilde{j}} \, \mathrm{d}t\,,
\label{eq:q_ast_2}
\end{equation}
where $\rho_{\text{p}}(t)=\ket{\psi_{\text{p}}(t)}\bra{\psi_{\text{p}}(t)}$. A discrete quantum-walk PageRank was proposed in Ref.~\cite{paparo2012google} and CTQW and quantum-stochastic-walk (QSW) versions of PageRank were proposed in Refs.~\cite{loke2017comparing,tang2020tensorflow}.
\subsection{Random-walk betweenness centrality}
\label{sec:rw_betweenness}
The betweenness centrality~\cite{newman2018networks} of a node quantifies the extent to which it lies on short paths that connect other nodes. To formulate classical and quantum random-walk-based betweenness centralities~\cite{newman2005measure}, we denote the transition tensor components with absorbing node $\ell$ in all layers by~\cite{sole2016random}
\begin{equation}
    {(T_{\ell})}^{i \alpha}_{j \beta} =
\begin{cases}
0\,,\quad &\text{if $i=k$}\\
T^{i \alpha}_{j \beta}\,,\quad &\text{if $i\neq \ell$}\,.
\end{cases}
\label{eq:transition_rw_betweenness}
\end{equation}
Alternatively, one can define an absorbing transition tensor in terms of node-layer tuples $(\ell,\gamma)$ and corresponding transition probabilities ${(T_{(\ell,\gamma)})}^{i \alpha}_{j \beta}$ as in Eq.~\eqref{eq:transition_rw_betweenness}.\footnote{Centralities that are based on nodes and centralities that are based on node-layers are relevant for different problems. It is important to consider the scientific question that one is asking. For example, one can ask if a person is important on social media (and hence consider a node) or alternatively if a person has an important Twitter account (and hence consider a node-layer).} Given $\mathcal{M}$ realizations of a classical or quantum random walk, the ensemble average of the number of times that a random walk that starts at node-layer $(o, \sigma)$ with destination $\ell$ passes through $(j,\beta)$ at time $t$ is
\begin{equation}
    (\tau_{\ell})_{j \beta}^{o \sigma} =  \lim_{\mathcal{M} \rightarrow \infty} \frac{1}{\mathcal{M}} \sum_{m=1}^\mathcal{M} \int_{0}^\infty z_{j \beta }^{o \sigma} (t,m) \, \mathrm{d}t\,,
\end{equation}
where $z_{j \beta }^{o \sigma} (t,m) \, \mathrm{d}t= 1$ if the random-walk realization $m\in\{1,\dots,\mathcal{M}\}$ is at $(j, \beta)$ at time $[t,t+\mathrm{d}t)$ and $z_{j \beta }^{o \sigma} (t,m) \, \mathrm{d}t= 0$ otherwise. 
Using
\begin{equation}
    \bar{z}_{j \beta }^{o \sigma} (t) = \frac{1}{\mathcal{M}} \sum_{m=1}^\mathcal{M} z_{j \beta }^{o \sigma} (t,m)
\label{eq:z_bar}
\end{equation}
yields
\begin{equation}
(\tau_{\ell})_{j \beta}^{o \sigma} = \int_0^\infty \bar{z}_{j \beta }^{o \sigma} (t) \, \mathrm{d}t\,.
\label{eq:tau_k}
\end{equation}
Note that $(\tau_{\ell})_{j \beta}^{o \sigma}$ diverges for $j=\ell$ because $\bar{z}_{j \beta }^{o \sigma} = 1$ after a walker hits the absorbing node $\ell$ at some finite time.\footnote{Although such a singularity also occurs in the discrete-time formulation of random-walk betweenness centrality, this was not pointed out in~\cite{sole2016random}.}
Averaging over all possible starting layers $\sigma$ and adding the number of times that a walk passes through node $j$ in any of the $K$ layers gives
\begin{equation}
    (\tau_\ell)^{o}_j = \frac{1}{K} (\tau_\ell)^{ o \sigma}_{j \beta} u^{\beta} u_{\sigma}\,,
    \label{eq:average1}
\end{equation}
where all components of the contravariant and covariant order-1 tensors $u^{\beta}$ and $u_{\sigma}$ are equal to $1$. After averaging over all possible origins $o$ and destinations $\ell$, the random-walk betweenness centrality of node $j$ is
\begin{equation}
    \tau_j = \frac{1}{N(N-1)} \sum_{\ell=1}^N {(\tau_{\ell})}^{o}_j u_o\,.
\label{eq:betweenness}
\end{equation}

We use $\tau_j^{\rm c}$ to denote the classical random-walk betweenness centrality of node $j$ and $\tau_j^{\rm q}$ to denote the quantum random-walk betweenness centrality of node $j$. To determine the occupation probability $\bar{z}_{j \beta }^{o \sigma} (t)$ for CRW and CTQW dynamics with absorbing transition-tensor components ${(T_{\ell})}^{i \alpha}_{j \beta}$, we modify the classical and quantum evolution equations \eqref{eq:cl_evolution} and \eqref{eq:q_evolution}. Reformulating Eq.~\eqref{eq:rw2} for an absorbing walk yields 
\begin{equation}
    \frac{\mathrm{d}}{\mathrm{d}t}p_{j \beta}(t) 
    =
    -\begin{cases}-\mathcal{A}^{\ell \gamma}_{j \beta} {D_{\ell \gamma}^{i \alpha}}^{-1} p_{i \alpha}(t) \,,~&\text{if $j=\ell$}\\
    \delta^{i \alpha}_{j \beta} p_{i \alpha}(t)-\mathcal{A}^{\ell \gamma}_{j \beta} {D_{\ell \gamma}^{i \alpha}}^{-1} p_{i \alpha}(t) \,,~&\text{if $j\neq \ell$}\\
    0^{i \alpha}_{j \beta} p_{i \alpha}(t)\,,\quad &\text{if $i=\ell$}\,,
    \end{cases}
\label{eq:rw_absorbing}
\end{equation}
where $0^{i \alpha}_{j \beta}$ is the $(2,2)$ tensor with entries that are equal to $0$. 

For our subsequent numerical calculations, we work in flattened notation and write the corresponding components of the classical absorbing-walk Hamiltonian with absorbing node-layer $\tilde{\ell}$ as 
\begin{align}
    \left[(\mathcal{H}_{\tilde{\ell}})_{\rm c}^{\rm a}\right]_{\tilde{i}\tilde{j}} =
\begin{cases}
(\mathcal{H}_{\rm c})_{\tilde{i}\tilde{j}}\,,\quad &\text{if $\tilde{j} \neq \tilde{\ell}$}\\
0\,,\quad &\text{if $\tilde{j} = \tilde{\ell}$}\,.
\end{cases}
\label{eq:H_c_absorbing}
\end{align}
Recall that, in flattened notation, the matrix element $-(\mathcal{H}_{\rm c})_{\tilde{i}\tilde{j}}$ (with $\tilde{i} \neq \tilde{j}$) describes the movement of a classical random walker from node-layer $\tilde{j}$ to node-layer $\tilde{i}$. We set $\left[(\mathcal{H}_{\tilde{\ell}})_{\rm c}^{\rm a}\right]_{\tilde{i}\tilde{j}}=0$ for $\tilde{j}=\tilde{\ell}$ because a random walker that reaches an absorbing node-layer cannot leave it anymore.

Analogously to the definition of the absorbing classical Hamiltonian \eqref{eq:H_c_absorbing}, we write the absorbing quantum Hamiltonian with absorbing node-layer $\tilde{\ell}$ as
\begin{align}
    \left[(\mathcal{H}_{\tilde{\ell}})_{\rm q}^{\rm a}\right]_{\tilde{i}\tilde{j}} =
\begin{cases}
(\mathcal{H}_{\rm q})_{\tilde{i}\tilde{j}}\,,\quad &\text{if $\tilde{j} \neq \tilde{\ell}$}\\
0\,,\quad &\text{if $\tilde{j} = \tilde{\ell}$}\,.
\end{cases}
\label{eq:H_q_absorbing}
\end{align}
Because of the absorbing nature of this random walk, $(\mathcal{H}_{\tilde{\ell}})_{\rm q}^{\rm a}$ is not Hermitian. Continuous-time quantum walks with non-Hermitian Hamiltonians are used as models of excitations that decay radiatively or via exciton recombination. See Refs.~\cite{mulken2007survival,mulken2011continuous,yalouz2018continuous} for examples of applications of non-Hermitian Hamiltonians to systems with absorption. One application of such models is as a phenomenological description of how excitations in light-harvesting systems propagate until they become trapped in reaction centers (\ie, absorbing nodes)~\cite{yalouz2018continuous}.

To determine the classical ($\tau^{\rm c}$) and quantum ($\tau^{\rm q}$) betweenness centralities using the Hamiltonians \eqref{eq:H_c_absorbing} (for the classical case) and \eqref{eq:H_q_absorbing} (for the quantum case), we employ a uniform walker distribution as initial distributions for $\mathbf{p}(0)$ and $\ket{\psi(0)}$, respectively. Using a uniformly-distributed initial walker configuration allows us to determine the evolution of absorbing CRWs and CTQWs for different initial node-layer configurations in parallel. This reduces the computational effort that is necessary to take a mean over all $\mathcal{N}$ initial walker positions. (See the summation over $\sigma$ and $o$ in Eqs.~\eqref{eq:average1} and \eqref{eq:betweenness}.) 

We now briefly summarize the steps that are necessary to compute the classical betweenness centrality $\tau^{\rm c}$. In flattened notation, we identify $\bar{z}$ with $e^{-(\mathcal{H}_{\tilde{\ell}})_{\rm c}^{\rm a} t} \mathbf{p}(0)$ and reformulate Eqs.~\eqref{eq:tau_k}--\eqref{eq:betweenness} as
\begin{equation}
\begin{split}
    \boldsymbol{\tau}^{\rm c} &=  \lim_{s\rightarrow 0} \frac{1}{\mathcal{N}(\mathcal{N}-1)}\sum_{\tilde{\ell}}\int_0^\infty e^{-s t} e^{-(\mathcal{H}_{\tilde{\ell}})_{\rm c}^{\rm a} t} \mathbf{p}(0)\, \mathrm{d}t\\
    &= \lim_{s\rightarrow 0}\frac{1}{\mathcal{N}(\mathcal{N}-1)}\sum_{\tilde{\ell}}\left[{s\mathds{1}+(\mathcal{H}_{\tilde{\ell}})_{\rm c}^{\rm a}}\right]^{-1} \mathbf{p}(0)\,,
\end{split}
\label{eq:betweenness_cl}
\end{equation}
where we have inserted the prefactor $e^{-s t}$ (with $s>0$) because $(\mathcal{H}_{\tilde{\ell}})_{\rm c}^{\rm a}$ is a singular matrix and one cannot solve the integral in Eq.~\eqref{eq:tau_k} in closed form for general networks. This formalism also avoids the divergence of the integral in Eq.~\eqref{eq:tau_k} that arises from the random-walk dynamics with an absorbing state. The sum in Eq.~\eqref{eq:betweenness_cl} is over all node-layers $\tilde{\ell}$, and the integral over $e^{-s t} e^{-(\mathcal{H}_{\tilde{\ell}})_{\rm c}^{\rm a} t}$ yields the resolvent $\left[{s\mathds{1}+(\mathcal{H}_{\tilde{\ell}})_{\rm c}^{\rm a}}\right]^{-1}$~\cite{pazy2012semigroups}.\footnote{Observe the similarity between the continuous-time formulation of random-walk betweenness centrality in \eqref{eq:betweenness_cl} and the discrete-time random-walk formulation of betweenness centrality in Ref.~\cite{sole2016random}.} See Ref.~\cite{wald2020classical} for further details on the resolvent formalism and its application to CRWs and CTQWs. Similar regularization approaches are common in scattering theory and quantum mechanics~\cite{thaller2005advanced,teschl2009mathematical}. For absorbing random walks with multiple absorbing nodes and different absorption rates, a generalized inverse of the combinatorial graph Laplacian was introduced in Ref.~\cite{jacobsen2018generalized}.
\begin{figure}
    \centering
    \includegraphics{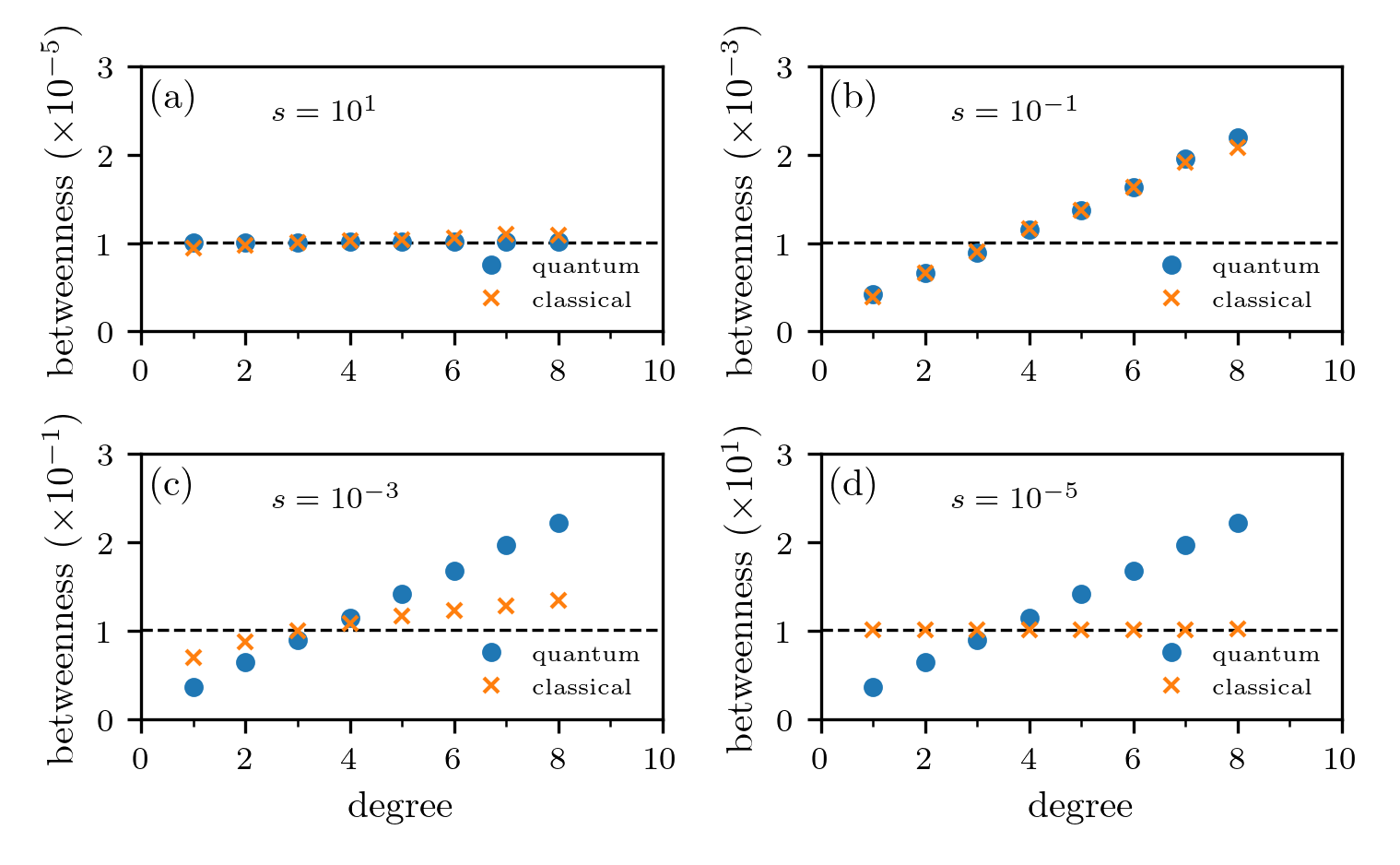}
    \caption{\textbf{Influence of the regularization term on classical and quantum betweenness centralities.} We show classical (orange crosses) and quantum (blue disks) betweenness centralities for (a) $s=10^{1}$, (b) $s=10^{-1}$, (c) $s=10^{-3}$, and (d) $s=10^{-5}$ in the regularization term of Eqs.~\eqref{eq:betweenness_cl} and \eqref{eq:qw_betweenness3} for a multilayer network with two ER layers and $N=50$ nodes in each layer. Interlayer edges connect each node-layer with its counterpart in the other layer. The expected mean degree in one layer is 2, and the expected mean degree in the other layer is 3. The dashed black line indicates the value $[s \mathcal{N}(\mathcal{N}-1)]^{-1}\approx 10^{-4}/s$. The product $\mathcal{N}(\mathcal{N}-1)$ is approximately $10^{-4}$, because the total number of node-layers in the multilayer network is $\mathcal{N}=100$. As the initial condition, we use a uniform distribution over all nodes.}
    \label{fig:s_variation}
\end{figure}

For quantum betweenness $\tau^{\rm q}$, one needs to compute the node-occupation probabilities in Eq.~\eqref{eq:tau_k} in terms of the squares of corresponding wave-function entries. For the quantum betweenness centrality, we thus obtain
\begin{equation}
    \begin{split}
\tau^{\rm q}_{\tilde{j}} &= \frac{1}{\mathcal{N}(\mathcal{N}-1)} \sum_{\tilde{\ell}} \int_0^\infty \braket{\tilde{j}|\rho_{\tilde{\ell}}(t)|\tilde{j}}\, \mathrm{d}t\,,
    \end{split}
\label{eq:qw_betweenness}
\end{equation}
where $\rho_{\tilde{\ell}}(t)= \ket{\psi_{\tilde{\ell}}(t)} \bra{\psi_{\tilde{\ell}}(t)}$ is the density matrix of the absorbing QW with absorbing node-layer $\tilde{\ell}$. We use $\ket{\psi_{\tilde{\ell}}(t)}=e^{-\iu (\mathcal{H}_{\tilde{\ell}})_{\rm q}^{\rm a} t} \ket{\psi(0)}$ and rewrite Eq.~\eqref{eq:qw_betweenness} as
\begin{equation}
    \begin{split}
\tau^{\rm q}_{\tilde{j}} &= \frac{1}{\mathcal{N}(\mathcal{N}-1)} \sum_{\tilde{\ell}} \int_0^\infty \sum_{l,m} e^{-\iu(\lambda_l^{(\tilde{\ell})}-\lambda_m^{(\tilde{\ell})})t}  \braket{e_l^{(\tilde{\ell})}|\psi(0)}  \braket{\psi(0)|e_m^{(\tilde{\ell})}} \braket{\tilde{j}|e_l^{(\tilde{\ell})}} \braket{e_m^{(\tilde{\ell})}|\tilde{j}}\, \mathrm{d}t\,,
    \end{split}
\label{eq:qw_betweenness2}
\end{equation}
where $e_l^{(\tilde{\ell})}$ and $\lambda_l^{(\tilde{\ell})}$, respectively, are the eigenvectors and corresponding eigenvalues of $(\mathcal{H}_{\tilde{\ell}})_{\rm q}^{\rm a}$. That is,
\begin{equation}
    (\mathcal{H}_{\tilde{\ell}})_{\rm q}^{\rm a} e_m^{(\tilde{\ell})} = \lambda_m^{(\tilde{\ell})} e_m^{(\tilde{\ell})}\,.
\end{equation}
The integral in Eq.~\eqref{eq:qw_betweenness2} diverges if $\lambda_m^{(\tilde{\ell})}-\lambda_n^{(\tilde{\ell})}=0$. Therefore, as with the classical random-walk betweenness centrality \eqref{eq:betweenness_cl}, we incorporate an additional prefactor $e^{-s t}$ (with $s > 0$) and obtain
\begin{equation}
    \tau^{\rm q}_{\tilde{j}} = \lim_{s \rightarrow 0}\frac{1}{\mathcal{N}(\mathcal{N}-1)} \sum_{\tilde{k}} \sum_{m,n} \frac{\braket{e_m^{(\tilde{\ell})}|\psi(0)}  \braket{\psi(0)|e_n^{(\tilde{\ell})}}}{s+\iu(\lambda_m^{(\tilde{\ell})}-\lambda_n^{(\tilde{\ell})})}\braket{\tilde{j}|e_m^{(\tilde{\ell})}} \braket{e_n^{(\tilde{\ell})}|\tilde{j}}\,.
\label{eq:qw_betweenness3}
\end{equation}
Observe the similarity in the mathematical structures of Eqs.~\eqref{eq:qw_betweenness2} and \eqref{eq:qw_betweenness3}. In the quantum case \eqref{eq:qw_betweenness3}, product states emerge as a result of mixing wave-function components.

For our numerical calculations in Sec.~\ref{sec:numerics}, we compute the classical \eqref{eq:betweenness_cl} and quantum \eqref{eq:qw_betweenness3} random-walk betweenness centralities with a value of $s$ that is small enough so that the ``damping'' term $e^{-s t}$ is not the dominant mechanism in the evolution of the classical and quantum absorbing random walks. As we detail in the next paragraph, the value of $s$ also needs to be large enough so that classical walkers do not get trapped in absorbing states. 

To study the influence of different values of $s$ on continuous-time random-walk betweenness centrality, we calculate $\tau^{\rm c}$ (see Eq.~\eqref{eq:betweenness_cl}) and $\tau^{\rm q}$ (see Eq.~\eqref{eq:qw_betweenness3}) for $s=10^{1}$, $s = 10^{-1}$, $s = 10^{-3}$, and $10^{-5}$ and a multilayer network with two Erd\H{o}s--R\'{e}nyi (ER) layers and 50 nodes in each layer (see Fig.~\ref{fig:s_variation}). The total number of node-layers is thus $\mathcal{N}=100$. As initial conditions, we use a uniform distribution over all nodes. That is, $\mathbf{p}(0)=\mathcal{N}^{-1}(1,1,\dots,1)^\top$ and $\ket{\psi(0)}=\mathcal{N}^{-1/2}(1,1,\dots,1)^\top$. For these initial distributions and in the limit $s\rightarrow\infty$ (\ie, with ``strong damping''), $\tau_j^{\rm q}$ and the components of $\boldsymbol{\tau}^{\rm c}$ approach $[s\mathcal{N}(\mathcal{N}-1)]^{-1}$. In Fig.~\ref{fig:s_variation}, we indicate this limiting value with a dashed black line. For $s=10^1$, we observe that the values of classical and quantum betweenness centralities are very close to the limiting value [see Fig.~\ref{fig:s_variation}(a)]. For $s=10^{-1}$, the values of both centrality measures increase with the node degree [see Fig.~\ref{fig:s_variation}(b)]. For progressively smaller values of $s$, classical random-walk betweenness centrality again approaches the limiting value $[s\mathcal{N}(\mathcal{N}-1)]^{-1}$ [see Fig.~\ref{fig:s_variation}(c,d)] because the classical walks get trapped in absorbing states. For small values of $s$, quantum random-walk betweenness centrality is affected mainly by the magnitude of $s$ because the mixing of the wave-function components in Eq.~\eqref{eq:qw_betweenness3} suppresses the effect of absorbing states. In the discussed example, a value of $s=10^{-1}$ is useful for calculating betweenness centrality. For the chosen initial conditions, substantially larger values of $s$ yield strong damping. This leads to the same, uniform node-occupation statistics in classical and quantum walks. If $s$ is substantially smaller than $10^{-1}$, the trapping of classical walks in absorbing states also leads to a uniform node-occupation distribution that one cannot use to distinguish between different nodes.
\subsection{Random-walk closeness centrality}
The closeness centrality of a node is based on the mean distance between that node and other nodes. One can compute some types of closeness centrality using absorbing random walks~\cite{newman2018networks}. The probability that a random walker reaches the absorbing node $\ell$ at time $h \leq t$ is~\cite{sole2016random}
\begin{equation}
    (q_\ell)^{o \sigma}(h \leq t) = u^{o \sigma} - \bar{z}_{j \beta }^{o \sigma} (t)u^{j \beta}\,,
\end{equation}
where $ \bar{z}_{j \beta }^{o \sigma} (t)$ is as defined in Eq.~\eqref{eq:z_bar}. The probability that the first-passage time occurs in the interval $[t,t+\mathrm{d}t)$ is
\begin{align}
\begin{split}
    (q_\ell)^{o \sigma}(h=t)\mathrm{d}t&=\lim_{\Delta t \rightarrow 0} \left[(q_\ell)^{o \sigma}(h \leq t+\Delta t)-(q_\ell)^{o \sigma}(h \leq t)\right]\\
&=-\lim_{\Delta t \rightarrow 0} \left[\bar{z}_{j \beta }^{o \sigma} (t+\Delta t)u^{j \beta}-\bar{z}_{j \beta }^{o \sigma} (t)u^{j \beta}\right]\\
&=-\lim_{\Delta t \rightarrow 0} \left[\frac{\bar{z}_{j \beta }^{o \sigma} (t+\Delta t)-\bar{z}_{j \beta }^{o \sigma} (t)}{\Delta t}\right] u^{j \beta} \Delta t\\
&= -\bar{z}_{j \beta }^{\prime o \sigma} (t)u^{j \beta} \mathrm{d}t\,,
    \end{split}
\end{align}
where $\bar{z}_{j \beta }^{\prime o \sigma} (t)$ is the first derivative of $\bar{z}_{j \beta }^{o \sigma} (t)$ with respect to time. After determining $(q_\ell)^{o \sigma}(h=t) \mathrm{d}t$, we compute the mean first-passage time of a random walker that starts at node-layer $(o,\sigma)$ and stops after reaching node $\ell$ in any layer. We obtain
\begin{align}
    \begin{split}
(H_\ell)^{o \sigma} &= \int_0^\infty t (q_\ell)^{o \sigma} (h=t)\,\mathrm{d}t =-\int_0^\infty t \bar{z}_{j \beta }^{\prime o \sigma} (t)u^{j \beta} \mathrm{d}t\\
&= -t \bar{z}_{j \beta }^{o \sigma} (t)u^{j \beta} \big|_0^\infty+\int_0^\infty  \bar{z}_{j \beta }^{o \sigma} (t)u^{j \beta} \mathrm{d}t \\
&= -\lim_{t \rightarrow \infty} t \bar{z}_{j \beta }^{o \sigma}(t)u^{j \beta}+\int_0^\infty  \bar{z}_{j \beta }^{o \sigma} (t)u^{j \beta} \mathrm{d}t \\
&= (\tau_{\ell})_{j \beta}^{o \sigma} u^{j \beta} \,.
\end{split}
\end{align}
In the last equality, we used the fact that $\bar{z}_{j \beta }^{o \sigma}(t)$ vanishes for long times because every walker is eventually absorbed by node $\ell$. Consequently, the mean first-passage time $(H_\ell)^{o \sigma}$ is equal to $(\tau_{\ell})_{j \beta}^{o \sigma} u^{j \beta}$ and is thus related to random-walk betweenness centrality (see Eq.~\eqref{eq:tau_k}). For discrete-time classical random walks, this relation was shown in Ref.~\cite{sole2016random}.

Averaging over $(H_\ell)^{o \sigma}$ for all nodes and layers yields the mean first-passage time
\begin{align}
\begin{split}
    h_{\ell} &= \frac{1}{(N-1) K} (H_\ell)^{o \sigma} u_{o \sigma} + \frac{1}{N} \pi_\ell^{-1}\\
    &=\frac{1}{(N-1) K} (\tau_{\ell})_{j \beta}^{o \sigma} u^{j \beta} u_{o \sigma} + \frac{1}{N} \pi_\ell^{-1}\,,
    \label{eq:closeness}
\end{split}
\end{align}
where $\pi_\ell$ is the (classical or quantum) random-walk occupation probability of node $\ell$. Because of the absorbing nature of the random walk that underlies the definition of $(\tau_{\ell})_{j \beta}^{o \sigma}$, we explicitly include the mean return time $\pi_\ell^{-1}$ in Eq.~\eqref{eq:closeness}. The random-walk closeness centrality of node $\ell$ is $1/h_{\ell}$. Equation \eqref{eq:closeness} thus connects closeness, betweenness, and occupation centralities.

As with random-walk betweenness centrality, the random-walk closeness centrality $h_{\ell}^{-1}$ is based on the occupation probability $\bar{z}_{j \beta }^{o \sigma} (t)$ of an absorbing random walk. We proceed as in Sec.~\ref{sec:rw_betweenness} and compute this probability, in flattened notation, using Eqs.~\eqref{eq:tau_k}, \eqref{eq:betweenness_cl}, and \eqref{eq:qw_betweenness3}. 
We write
\begin{align}
    h_{\tilde{\ell}} =\frac{1}{\mathcal{N}-1} (\tau_{\tilde{\ell}})_{\tilde{j}}^{\tilde{o}} u^{\tilde{j}} u_{\tilde{o}} + \frac{1}{\mathcal{N}} \pi_{\tilde{\ell}}^{-1}\,.
    \label{eq:closeness2}
\end{align}
In flattened notation, $\pi_{\tilde{\ell}}$ is given by $p_{\tilde{\ell}}^\ast$ in the classical case (see Eq.~\eqref{eq:p_ast}) and $q_{\tilde{\ell}}^\ast$ in the quantum case (see Eq.~\eqref{eq:q_ast_1}). In Table~\ref{tab:centralities}, we summarize the classical and quantum multilayer random-walk centrality measures that we have discussed.
\begin{landscape}
\begin{table}[p]
\centering
\renewcommand*{\arraystretch}{4.5}
\begin{tabular}{|c|*{2}{c|}}\hline
Centrality
& \makebox[8em]{Classical} & \makebox[8em]{Quantum} \\\hline\hline
occupation & \,\,\,  \makecell[l]{ $\mathbf{p}^*$, which is the stationary state of $\frac{\mathrm{d}}{\mathrm{d}t}{\mathbf{p}}=-\mathcal{H}_{\rm c} \mathbf{p}$ \\ with the CRW Hamiltonian $\mathcal{H}_{\rm c}=\mathbf{L}_{\rm M} \mathbf{D}_{\rm M}^{-1}$} \,\, & \,\,\, 
\makecell[l]{$\mathbf{q}^*$, which is the long-time mean \eqref{eq:q_ast_1} of the corresponding \\ components of the density matrix $\rho(t)=\ket{\psi(t)}\bra{\psi(t)}$  with  \\ $\frac{\mathrm{d}}{\mathrm{d}t}{\ket{\psi}}=- \iu \mathcal{H}_{\rm q} \ket{\psi}$ and the CTQW Hamiltonian $\mathcal{H}_{\rm q}=\hat{\mathbf{L}}_{\rm M}$} \,\, \\[8pt]\hline
PageRank & \,\,\,  \makecell[l]{$\mathbf{p}_{\rm p}^*$, which is based on the PageRank Hamiltonian \\ $\mathcal{H}_{\rm c}^{\text{p}} = a \mathcal{H}_{\rm c}+(1-a) (\mathds{1}-\mathcal{N}^{-1}\mathbf{U})$ with \\ teleportation probability $1-a$} \,\, & \,\,\, 
\makecell[l]{$\mathbf{q}_{\rm p}^*$, which is based on the PageRank Hamiltonian\\  $\mathcal{H}_{\rm c}^{\text{q}} = a\mathcal{H}_{\rm q}+(1-a)(\mathds{1}-\mathcal{N}^{-1}\mathbf{U})$ with \\ teleportation probability $1-a$} \,\, \\[8pt]\hline
betweenness &\,\,\, 
\makecell[l]{$\boldsymbol{\tau}^{\rm c} \propto \lim_{s\rightarrow 0}\sum_{\tilde{\ell}}\left[{s\mathds{1}+(\mathcal{H}_{\tilde{\ell}})_{\rm c}^{\rm a}}\right]^{-1} \mathbf{p}(0)$, where \\ $(\mathcal{H}_{\tilde{\ell}})_{\rm c}^{\rm a} $ is the absorbing-walk Hamiltonian \eqref{eq:H_c_absorbing}\\ with absorbing node-layer $\tilde{\ell}$} \,\, & \,\,\,
\makecell[l]{$\tau^{\rm q}_j \propto \sum_{\tilde{\ell}} \sum_{m,n} \frac{\braket{e_m^{({\tilde{\ell}})}|\psi(0)}  \braket{\psi(0)|e_n^{({\tilde{\ell}})}}}{s+\iu(\lambda_m^{({\tilde{\ell}})}-\lambda_n^{({\tilde{\ell}})})}\braket{{\tilde{j}}|e_m^{({\tilde{\ell}})}} \braket{e_n^{({\tilde{\ell}})}|{\tilde{j}}}$, where\\ $e_m^{({\tilde{\ell}})}$ and $\lambda_m^{({\tilde{\ell}})}$, respectively, are the eigenvectors and \\ eigenvalues of the absorbing-quantum-walk\\ Hamiltonian \eqref{eq:H_q_absorbing} with absorbing node-layer $\tilde{\ell}$} \,\, \\[20pt]\hline
closeness & \,\,\, \makecell[l]{$({h_{\tilde{\ell}}^{\rm c}})^{-1}$, where $h_{\tilde{\ell}}^{\rm c}=\frac{1}{\mathcal{N}-1} (\tau_{{\tilde{\ell}}}^{\rm c})_{\tilde{j}}^{\tilde{o}} u^{\tilde{j}} u_{\tilde{o}} + \frac{1}{\mathcal{N}} {p^*_{\tilde{\ell}}}^{-1}$\\ is the mean first-passage time} \,\, & 
 \,\,\, \makecell[l]{$({h_{\tilde{\ell}}^{\rm q}})^{-1}$, where $h_{\tilde{\ell}}^{\rm q}=\frac{1}{\mathcal{N}-1} (\tau_{{\tilde{\ell}}}^{\rm q})_{\tilde{j}}^{\tilde{o}} u^{\tilde{j}} u_{\tilde{o}} + \frac{1}{\mathcal{N}} {q^*_{\tilde{\ell}}}^{-1}$\\ is the mean first-passage time} \,\, \\[8pt]\hline
\end{tabular}
\vspace{1mm}
\caption{\textbf{Definitions of several classical and quantum multilayer random-walk centrality measures.} We summarize the employed definitions of classical and quantum random-walk occupation, PageRank, betweenness, and closeness centralities. The classical random-walk Hamiltonian is the product of the combinatorial supra-Laplacian $\mathbf{L}_{\rm M}$ and the inverse-degree supra-matrix $\mathbf{D}_{\rm M}^{-1}$. For the (continuous-time) quantum walk, we use the normalized supra-Laplacian matrix $\hat{\mathbf{L}}_{\rm M}$. The identity matrix is $\mathds{1}$, and $\mathbf{U}$ is the matrix of $1$ entries. We present all centralities in flattened multilayer notation (\ie, each vector element corresponds to a node-layer). The number of node-layers is $\mathcal{N}$. We use $u^{\tilde{j}}$ and $u_{\tilde{o}}$, respectively, to denote the components of the contravariant and covariant order-1 tensors that are equal to $1$.}
\label{tab:centralities}
\end{table}
\end{landscape}
\section{Numerical examples}
\label{sec:numerics}
\begin{figure}[htp]
    \centering
    \includegraphics{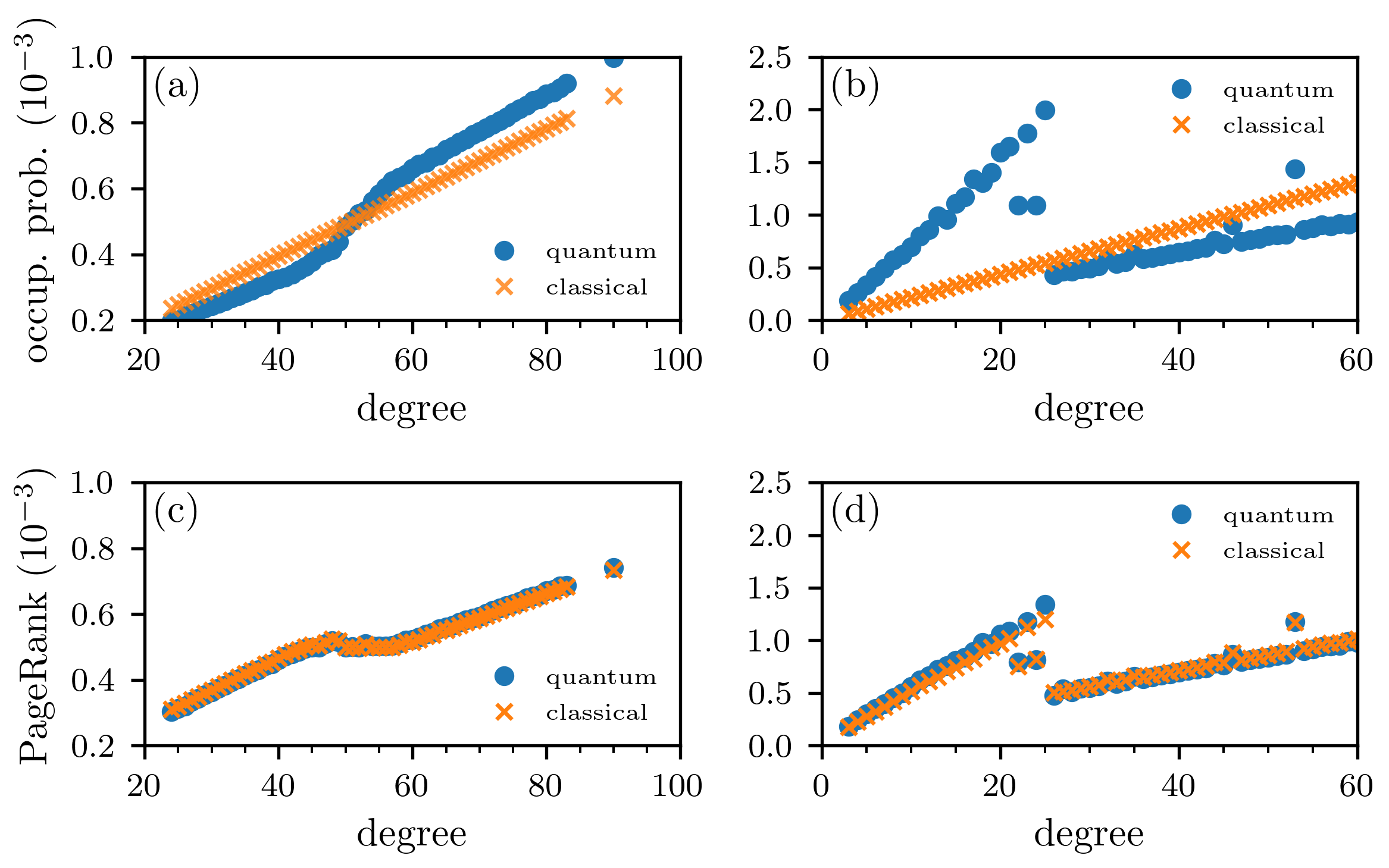}
    \caption{\textbf{Classical and quantum occupation centrality and PageRank on synthetic multilayer networks.} We show classical (orange crosses) and quantum (blue disks) random-walk centralities [with occupation centrality in panels (a,b) and PageRank centrality with $a=0.85$ in panels (c,d)] for two multilayer networks with two layers and $N=1000$ nodes in each layer. (a,c) The multilayer network consists of two ER layers and interlayer edges that connect each node-layer with its counterpart in the other layer. The expected mean degree in one layer is 40, and the expected mean degree in the other layer is 60. (b,d) The multilayer network consists of one ER layer and one BA layer. Interlayer edges connect each node-layer with its counterpart in the other layer. The expected mean degree of the ER layer is 40. In the BA layer, we start with a dyad and iteratively add new nodes until we reach $N=1000$ nodes. Each new node has 2 edges that connect to existing nodes using linear preferential attachment. As the initial condition for each calculation, we use a uniform distribution over all nodes.
    }
    \label{fig:random_networks_oc_pr_synthetic}
\end{figure}
In this section, we present some numerical examples of the random-walk centrality measures from Sec.~\ref{sec:centralities}. See Table~\ref{tab:centralities} for a summary of these centralities. We compare these classical and quantum random-walk centralities for two types of synthetic and empirical multilayer networks. 

Our first example of a synthetic multilayer network consists of two $G(N,p)$ ER layers with $N=1000$ nodes each and $p=0.04$ and $p=0.06$, respectively. The expected mean degree of one layer is 40, and the expected mean degree of the other layer is 60. 
Our second synthetic multilayer network consists of one $G(N,p)$ ER layer with $p=0.04$ and one Barab\'{a}si--Albert (BA) layer. The former has an expected mean degree of 40. To construct the latter, we start with one dyad and iteratively add new nodes until we reach $N=1000$ nodes.  Each new node has 2 edges that connect to existing nodes using linear preferential attachment~\cite{newman2018networks}. Both layers have $N=1000$ nodes. In both of these examples, each node-layer is adjacent to its counterpart in the other layer but not to any other node-layers in a different layer. Consequently, both synthetic multilayer networks are multiplex networks (see Sec.~\ref{sec:multilayer}).

We also calculate centralities for two empirical multilayer networks. The first one is the Lazega Law Firm network~\cite{lazega2001collegial,Lazega_data}, which has $N=71$ nodes and 2571 edges. It has three different edge types, which encode different relationships between partners and associates of a corporate law firm. The second empirical multilayer network is the London metropolitan (``Tube'') network~\cite{de2014navigability,London_data}, which has $N=369$ nodes and 441 edges. It also has three edge types, which encode connections within the three layers (Underground, Overground, and Docklands Light Railway) of London metro stations. As in the synthetic networks, in each of these examples, each node-layer is adjacent to its counterpart in the other layer (and not to any other node-layers in a different layer).
\begin{figure}
    \centering
    \includegraphics{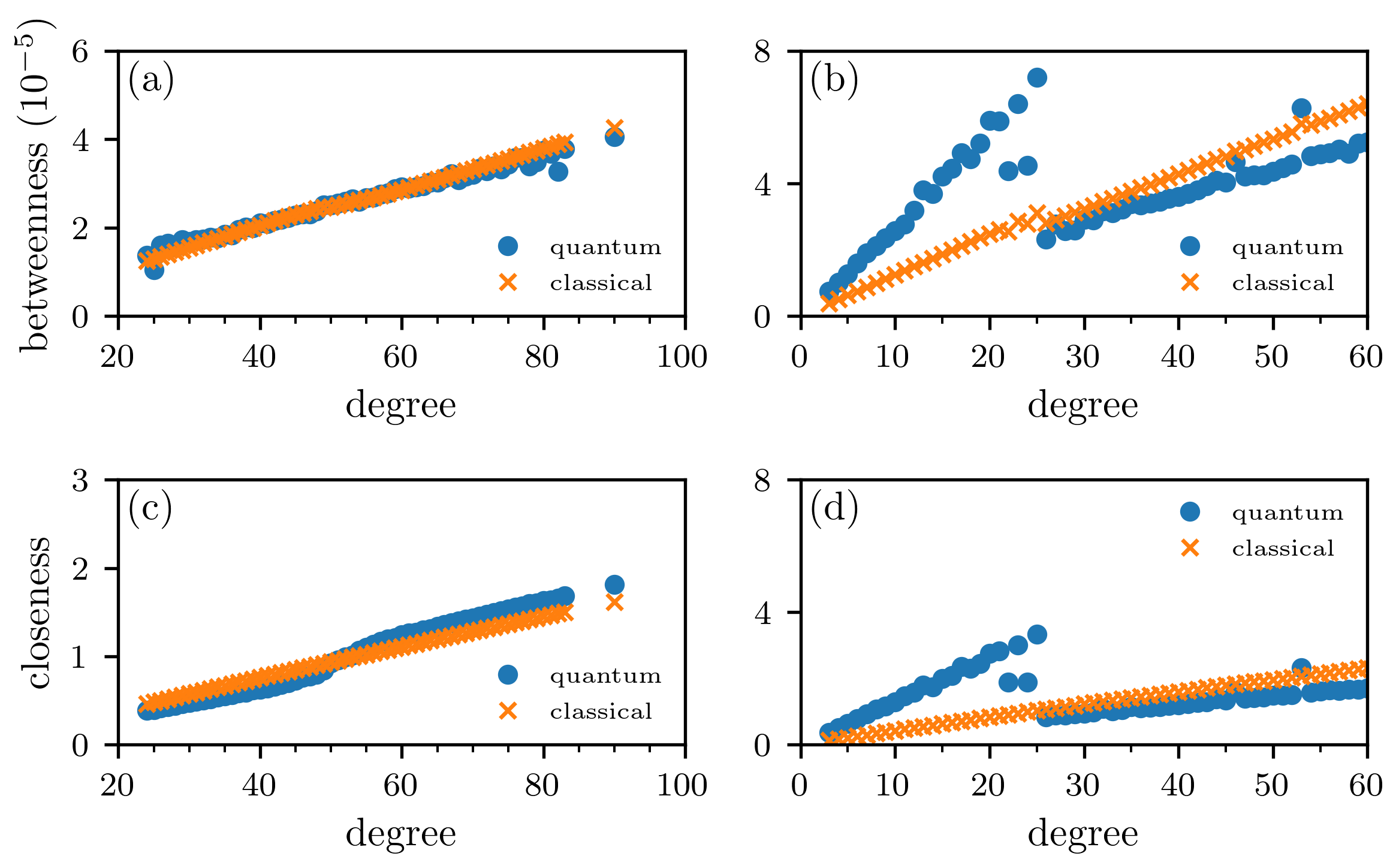}
    \caption{\textbf{Classical and quantum random-walk betweenness and random-walk closeness on synthetic multilayer networks.} We show the classical (orange crosses) and quantum (blue disks) random-walk centralities [with betweenness centrality in panels (a,b) and closeness centrality in panels (c,d)] for two multilayer networks with two layers and $N=1000$ nodes in each layer. To compute the resolvent (see Table~\ref{tab:centralities}), we set $s=0.01$. (a,c) The multilayer network consists of two ER layers and interlayer edges that connect each node-layer with its counterpart in the other layer. The expected mean degree in one layer is 40, and the expected mean degree in the other layer is 60. (b,d) The multilayer network consists of one ER layer and one BA layer. Interlayer edges connect each node-layer with its counterpart in the other layer. The expected mean degree of the ER layer is 40. In the BA layer, we start with a dyad and iteratively add new nodes until we reach $N=1000$ nodes. Each new node has 2 edges that connect to existing nodes using linear preferential attachment. As the initial condition for each calculation, we use a uniform distribution over all nodes.}
    \label{fig:random_networks_bc_cc_synthetic}
\end{figure}

To illustrate the results of our calculations, we plot node-layer centralities versus node-layer degrees. In Figs.~\ref{fig:random_networks_oc_pr_synthetic}(a,b), we show our results for classical and quantum random-walk occupation centralities on our synthetic multilayer networks. The linear dependence of the classical occupation centrality on node-layer degree that we observe in our numerical results is explained by the analytical result \eqref{eq:classical_analytic}. Unlike classical walks, quantum walks do not approach a stationary state and do not satisfy Eq.~\eqref{eq:classical_analytic}. Instead, their long-time behavior is characterized by the long-time mean \eqref{eq:q_ast_1}. Our results in Figs.~\ref{fig:random_networks_oc_pr_synthetic}(a,b) suggest that the node-layer occupation properties of CTQWs are not captured by node-layer degree alone (so, in particular, they are not proportional to node-layer degree). These properties also depend on other structural features of the underlying networks. The minimum and maximum degrees of the underlying BA layer in Figs.~\ref{fig:random_networks_oc_pr_synthetic}(b,d) are 2 and 52, respectively. In the ER layer, the minimum and maximum degrees are 21 and 60, respectively. In Fig.~\ref{fig:random_networks_oc_pr_synthetic}(d), we observe that some node-layers (specifically, those with degrees that are smaller than 21) in the BA layer have a larger classical and quantum PageRank than some node-layers in the ER layer, leading to deviations from the monotonic behavior of classical occupation centrality in Fig.~\ref{fig:random_networks_oc_pr_synthetic}(b).
\begin{figure}
    \centering
    \includegraphics{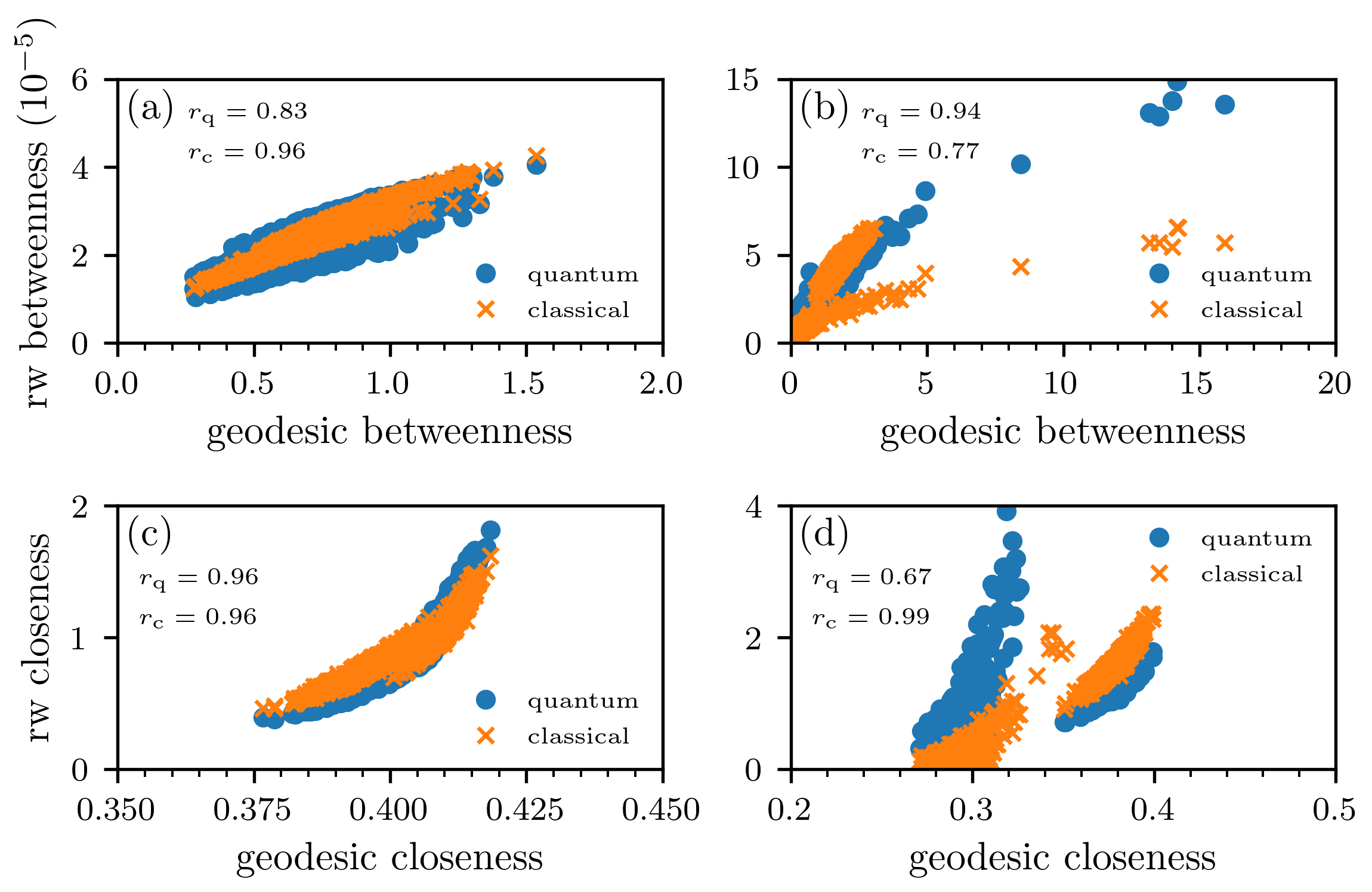}
    \caption{\textbf{Correlations between random-walk and geodesic centralities on synthetic multilayer networks.} We show the correlations between random-walk (``rw'') and geodesic betweenness and closeness on two multilayer networks with two layers and $N=1000$ nodes in each layer. In each panel, we use $r_{\rm c}$ and $r_{\rm q}$ to indicate the numerical values of the Pearson correlation coefficients for classical and quantum centralities, respectively. We indicate the classical and quantum random-walk centralities using orange crosses and blue disks, respectively. To compute the resolvent (see Table~\ref{tab:centralities}), we set $s=0.01$. (a,c) The multilayer network consists of two ER layers with interlayer edges that connect each node-layer with its counterpart in the other layer. The expected mean degree in one layer is 40, and the expected mean degree in the other layer is 60. (b,d) The multilayer network consists of one ER layer and one BA layer. Interlayer edges connect each node-layer with its counterpart in the other layer. The expected mean degree of the ER layer is 40. In the BA layer, we start with a dyad and iteratively add new nodes until we reach $N=1000$ nodes. Each new node has 2 edges that connect to existing nodes using linear preferential attachment. As the initial condition for each calculation, we use a uniform distribution over all nodes.}
    \label{fig:correlations_synthetic}
\end{figure}

In our two synthetic multilayer networks, we observe more pronounced differences between classical and quantum occupation centralities than was the case for the monolayer ER and BA networks that were studied in Ref.~\cite{faccin2013degree}. In agreement with previous results on node-occupation statistics of classical and quantum walks~\cite{faccin2013degree}, we observe that the differences between these two formulations of occupation centrality are larger in the synthetic multilayer network that includes a BA layer. This arises from the large degree heterogeneity in that layer~\cite{faccin2013degree}.

In Figs.~\ref{fig:random_networks_oc_pr_synthetic}(c,d), we show the occupation probabilities for classical and quantum PageRank with a teleportation probability of $1-a=0.15$. The classical and quantum occupation statistics are almost identical, which is very different from the case without teleportation in Figs.~\ref{fig:random_networks_oc_pr_synthetic}(a,b). This observation suggests that nonzero teleportation probabilities counteract differences in the node-occupation statistics between classical and quantum walks. In the limit $r\rightarrow 0$ (in which there is teleportation dynamics only), classical and quantum PageRank yield the same occupation statistics because all nodes are occupied with the same probability.

To further compare the classical random-walk and quantum random-walk centralities on multilayer networks, we also compute the random-walk betweenness and random-walk closeness for our two synthetic multilayer networks. In Figs.~\ref{fig:random_networks_bc_cc_synthetic}(a,b), we show our results for betweenness centrality and observe marked differences in the two synthetic multilayer networks. For the multilayer network with a BA layer, the quantum random-walk betweennesses of small-degree nodes is larger than their classical random-walk betweennesses. The opposite holds for most nodes with degrees that are at least 26 in Fig.~\ref{fig:random_networks_bc_cc_synthetic}(b). Our computation of closeness centralities in Fig.~\ref{fig:random_networks_bc_cc_synthetic}(c,d) illustrates that classical and quantum random-walk closeness centralities give node characterizations that are qualitatively similar to those of occupation centrality [see Figs.~\ref{fig:random_networks_oc_pr_synthetic}(a,b)]. This observation is intuitive because occupation centrality appears in the definition \eqref{eq:closeness2} of random-walk closeness centrality.
\begin{figure}
    \centering
    \includegraphics{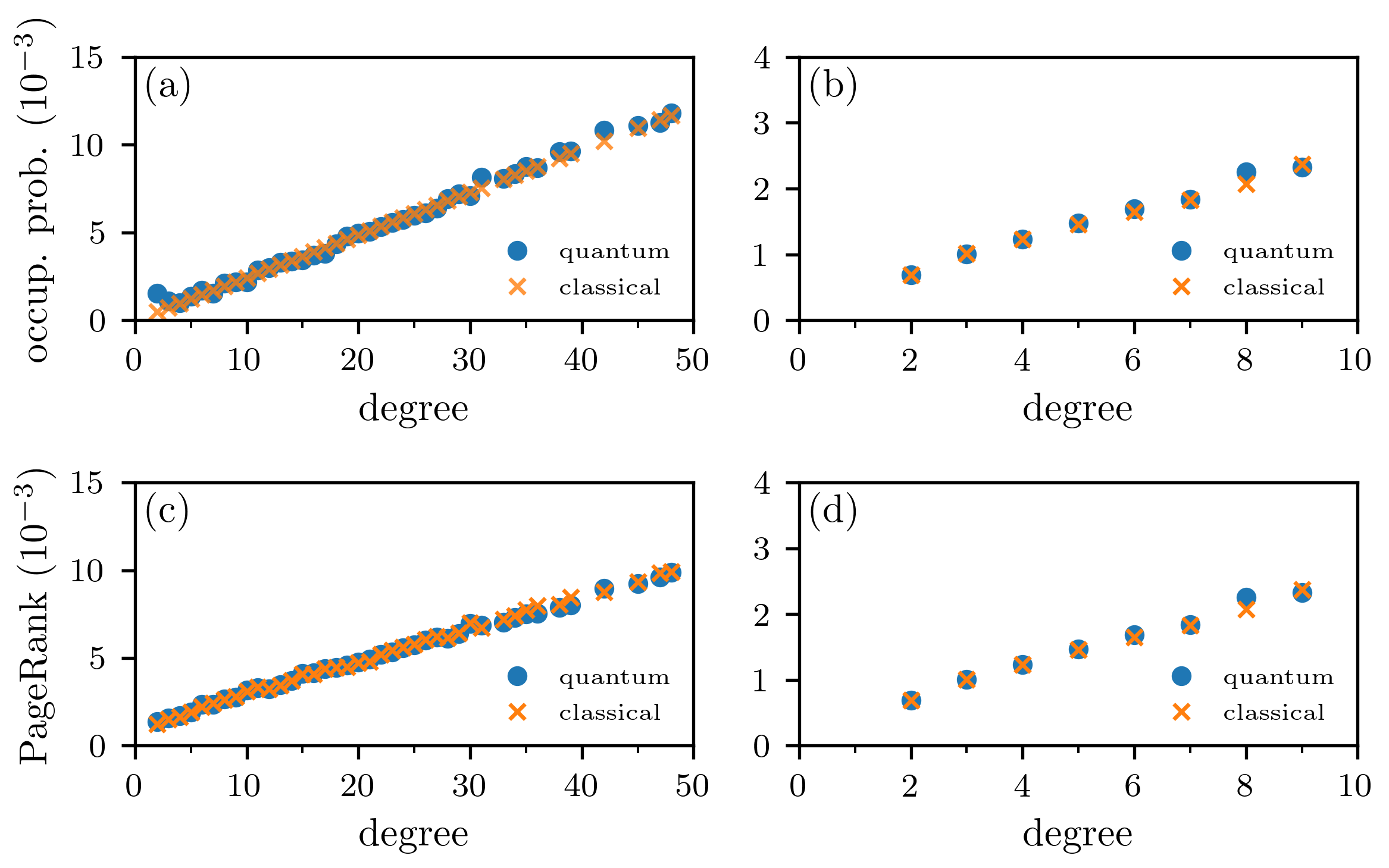}
    \caption{\textbf{Classical and quantum occupation centrality and PageRank on empirical multilayer networks.} We show classical (orange crosses) and quantum (blue disks) random-walk centralities [with occupation centrality in panels (a,b) and PageRank centrality for $a=0.85$ in panels (c,d)] for two empirical multilayer networks. (a,c) The Lazega Law Firm network, which has $N=71$ nodes and 2571 edges. It has three edge types, which represent different relationships between partners and associates of a corporate law firm. (b,d) A multilayer network of the London metropolitan (``Tube'') transportation network, which has $N=369$ nodes and 441 edges. It has three edge types, which encode connections within the three layers (Underground, Overground, and Docklands Light Railway) of London metro stations. In both multilayer networks, interlayer edges connect each node-layer with its counterpart in the other layer. As the initial condition for each calculation, we use a uniform distribution over all nodes.}
    \label{fig:empirical_networks_oc_pr_empirical}
\end{figure}
One can also calculate betweenness and closeness centralities that are based on shortest paths, rather than on random walks. Such geodesic centrality measures give an alternative notion of betweenness and closeness in networks. In Fig.~\ref{fig:correlations_synthetic}, we plot the correlations of random-walk-based and geodesic betweenness and closeness centralities for our synthetic multilayer networks. We observe that random-walk-based and geodesic centralities are positively correlated with each other. For the multilayer network with two ER layers, the Pearson correlation coefficients are 0.96 (classical) and 0.83 (quantum) for betweenness centrality and 0.96 (classical) and 0.96 (quantum) for closeness centrality. For the multilayer network with an ER layer and a BA layer, the Pearson correlation coefficients are 0.77 (classical) and 0.94 (quantum) for betweenness centrality and 0.99 (classical) and 0.67 (quantum) for closeness centrality. The p-values for all correlation coefficients are smaller than (single) machine precision.
\begin{figure}
    \centering
    \includegraphics{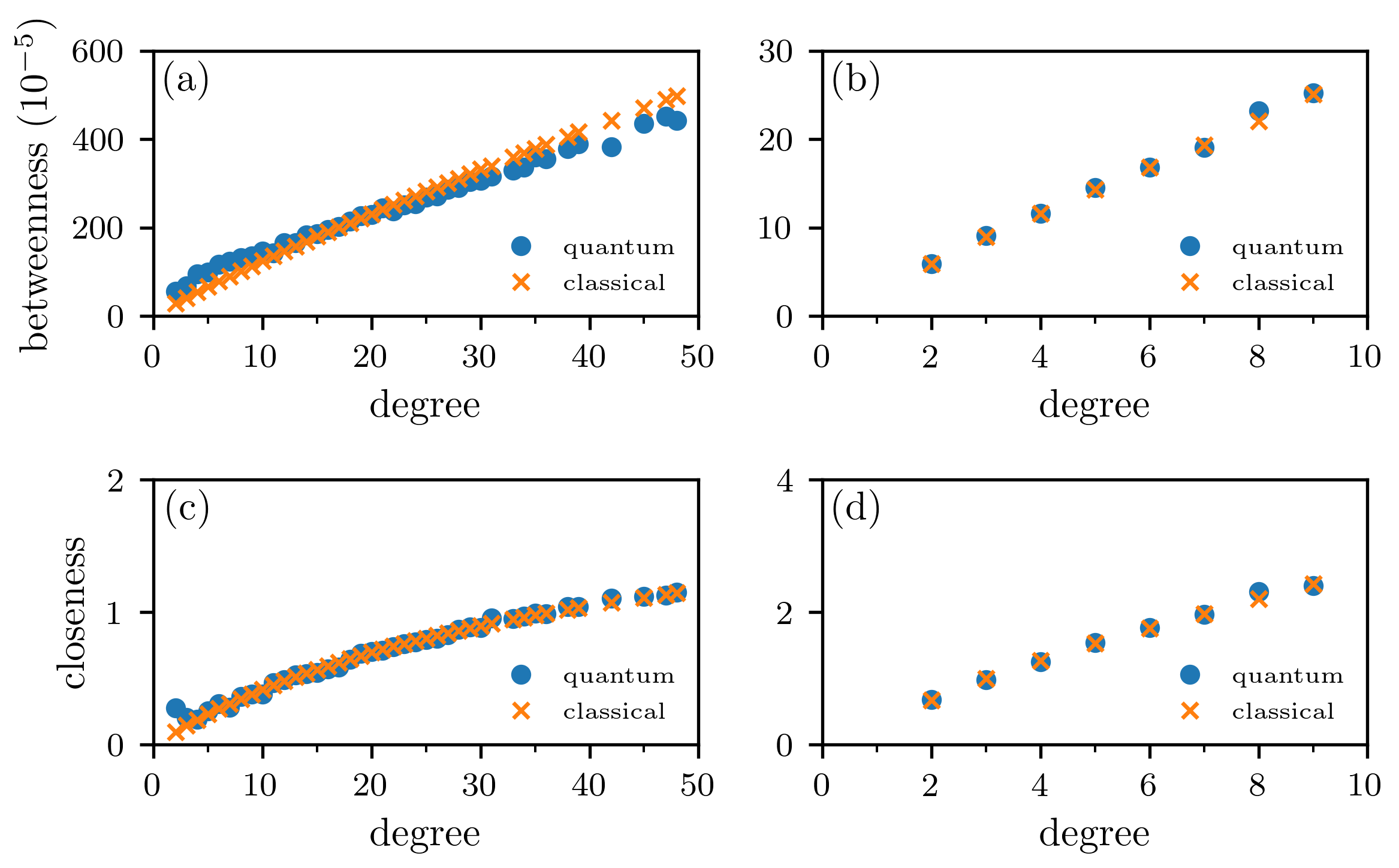}
    \caption{\textbf{Classical and quantum random-walk betweenness and random-walk closeness on empirical multilayer networks.} We show the classical (orange crosses) and quantum (blue disks) random-walk centralities [with betweenness centrality in panels (a,b) and closeness centrality in panels (c,d)] for two empirical multilayer networks. To compute the resolvent (see Table~\ref{tab:centralities}), we set $s=0.01$. (a,c) The Lazega Law Firm network, which has $N=71$ nodes and 2571 edges. It has three edge types, which encode different relationships between partners and associates of a corporate law firm. (b,d) A multilayer network of the London metropolitan (``Tube'') network, which has $N=369$ nodes and 441 edges. It has three edge types, which encode connections within the three layers (Underground, Overground, and Docklands Light Railway) of London metro stations. In both multilayer networks, interlayer edges connect each node-layer with its counterpart the other layer. As the initial condition for each calculation, we use a uniform distribution over all nodes.}
    \label{fig:random_networks_bc_cc_empirical}
\end{figure}

We now calculate classical and quantum occupation, PageRank, betweenness, and closeness centralities for our two empirical networks. In Fig.~\ref{fig:empirical_networks_oc_pr_empirical}, we show the results of our numerical calculations of occupation and PageRank centralities and observe that the differences between these classical and quantum centralities are smaller than those that we observed for the synthetic multilayer networks in Fig.~\ref{fig:random_networks_oc_pr_synthetic}. Similarly, the differences between classical and quantum betweenness and closeness centralities that we show in Fig.~\ref{fig:random_networks_bc_cc_empirical} are smaller than those that we observed for the synthetic multilayer networks in Fig.~\ref{fig:random_networks_bc_cc_synthetic}.

\begin{figure}
    \centering
    \includegraphics{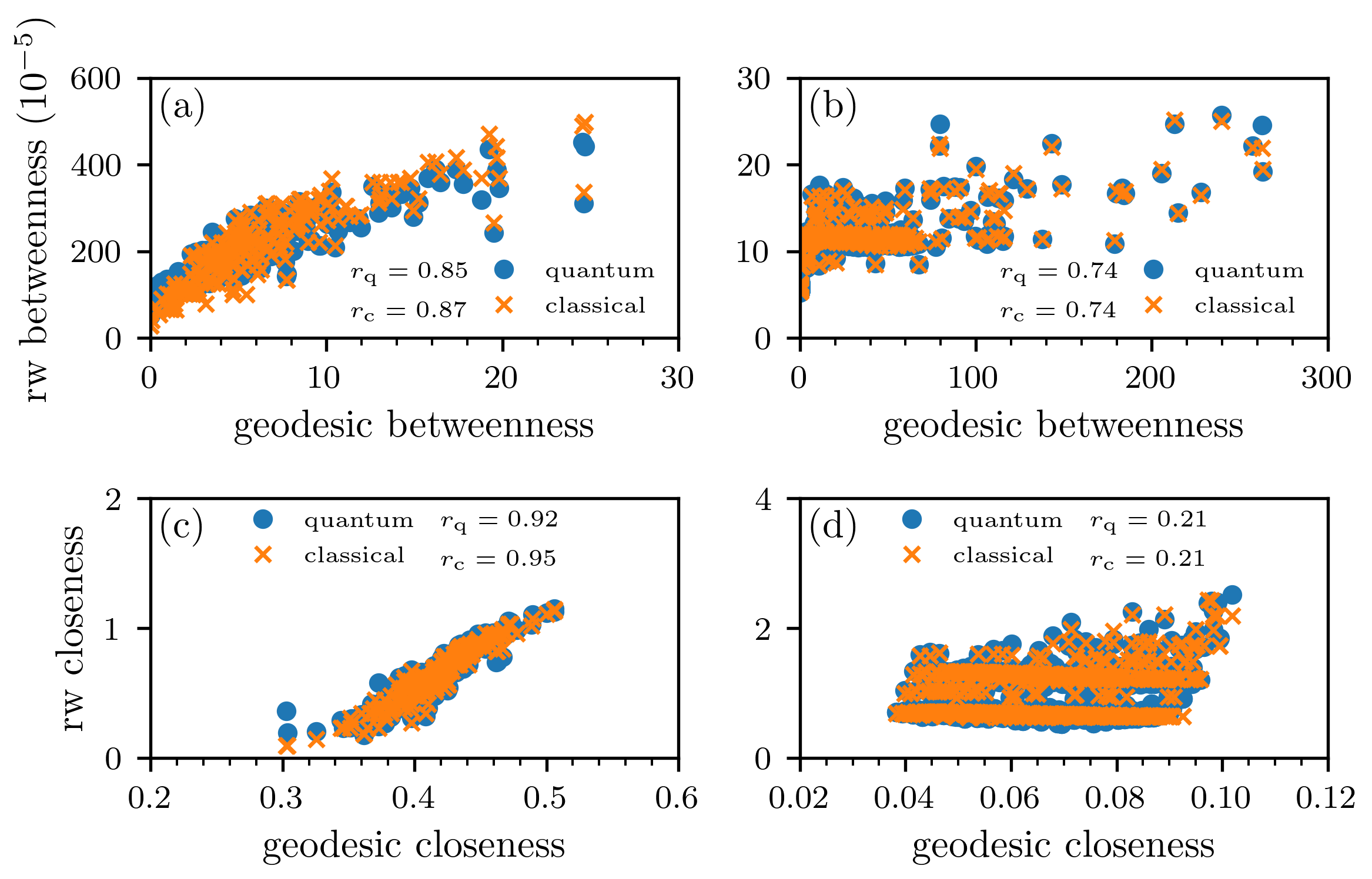}
    \caption{\textbf{Correlations between random-walk and geodesic centralities on empirical multilayer networks.} We show the correlations between random-walk (rw) and geodesic betweenness and closeness on two empirical multilayer networks. In each panel, we use $r_{\rm c}$ and $r_{\rm q}$ to indicate the numerical values of the Pearson correlation coefficients for classical and quantum centralities, respectively. We indicate the classical and quantum random-walk centralities using orange crosses and blue disks, respectively. To compute the resolvent (see Table~\ref{tab:centralities}), we set $s=0.01$. (a,c) The Lazega Law Firm network, which has $N=71$ nodes and 2571 edges. (b,d) A multilayer network of the London metropolitan (``Tube'') network with $N=369$ nodes and 441 edges. It has three edge types, which encode connections in the three layers (Underground, Overground, and Docklands Light Railway) of London metro stations. In both of these multilayer networks, interlayer edges connect each node-layer with its counterpart in the other layer. As the initial condition for each calculation, we use a uniform distribution over all nodes.}
    \label{fig:correlations_empirical}
\end{figure}

As with our two synthetic multilayer networks, we examine the correlations between geodesic betweenness and closeness and their random-walk counterparts (see Fig.~\ref{fig:correlations_empirical}). For the Lazega Law Firm network, the Pearson correlation coefficients are 0.87 (classical) and 0.85 (quantum) for betweenness centrality and 0.95 (classical) and 0.92 (quantum) for closeness centrality. For the London Tube transportation network, the Pearson correlation coefficients are 0.74 (classical) and 0.74 (quantum) for betweenness centrality and 0.21 (classical) and 0.21 (quantum) for closeness centrality. The p-values for all correlation coefficients are smaller than (single) machine precision.

We find that large differences between classical and quantum random-walk occupation centrality are associated with a large degree heterogeneity in the underlying network. These observations are similar to the results in Ref.~\cite{faccin2013degree}. However, our results for the multilayer network with a BA layer contrast starkly with those in the literature of which we are aware. Specifically, the difference between classical and quantum random-walk occupation centrality in this synthetic multilayer network is significantly larger than those that were reported for monolayer networks in Refs.~\cite{faccin2013degree,wald2020classical}.
\section{Conclusions and Discussion}
\label{sec:conclusion}
We formulated and analyzed classical and quantum continuous-time random-walk centrality measures in multilayer networks. We have three main contributions. First, we generalized the classical discrete-time random-walk centralities of Ref.~\cite{sole2016random} to continuous time. Second, we introduced and studied continuous-time quantum-walk (CTQW) generalizations of occupation, PageRank, betweenness, and closeness centralities. Third, we formulated our mathematical framework of continuous-time classical and quantum centrality measures in a way that makes it applicable to both monolayer and multilayer networks. Our results complement earlier studies that focused on quantum occupation~\cite{sanchez2012quantum,rossi2014node,izaac2017centrality,wald2020classical} and PageRank~\cite{loke2017comparing,tang2020tensorflow} centrality measures on monolayer networks.

In continuous time, the evolution of the underlying classical and quantum walks is described by different Hamiltonians, which we summarized in Table~\ref{tab:centralities}. One can derive other classical and quantum continous-time random-walk centralities by modifying these Hamiltonians and tracking different properties of absorbing random walks, which are what we used to define random-walk betweenness and closeness.

There are various interesting ways to build on our work. One worthwhile direction is to develop multilayer extensions of generalized versions of PageRank~\cite{gleich2015pagerank} with various teleportation strategies and to compare classical and quantum versions of these generalizations. One can use and adapt existing multilayer generalizations, such as a multilayer personalized PageRank of \cite{jeub2015local} (in which the teleportation strategy depends on the initial location of a random walker) and multilayer versions of PageRank that include both node teleportation and layer teleportation \cite{taylor2019tunable}. One can also generalize other versions of PageRank, such as ones with ``smart teleportation'' \cite{smart2012}, to multilayer networks and then compare classical and quantum versions of such generalizations. Another important research direction is to consider more general types of quantum states.
We considered pure quantum states in our derivations and numerical experiments, and it will be interesting to explore the effects of entangled and mixed states on the node-occupation properties of quantum walks. One can also use the framework of quantum stochastic walks (QSWs)~\cite{whitfield2010quantum,wald2020classical} to interpolate between classical and quantum walks, and it is worthwhile to study QSWs on multilayer networks.
\section*{Acknowledgements}

We thank Sascha Wald for helpful discussions.


\bibliographystyle{siamplain}
\bibliography{refs}
\end{document}